\documentclass[conference]{IEEEtran}

\usepackage{graphicx}
\usepackage{amsmath}
\usepackage{amssymb}
\newcommand{\var}[1]{\mbox{Var}\left(#1\right)}

\begin{document}

\title{Secret Key Agreement Over Multipath Channels Exploiting a Variable-Directional Antenna}
\author{\authorblockN{Valery~Korzhik, Viktor~Yakovlev}
\authorblockA{
Members of IEEE \\
State University of Telecommunication \\
St. Petersburg, Russia  \\
E-mail: \{korzhik1,viyak\}@bk.ru}
\and
\authorblockN{Guillermo Morales-Luna}
\authorblockA{Computer Science Department \\
CINVESTAV-IPN \\
Mexico City, Mexico \\
E-mail: gmorales@cs.cinvestav.mx}
\and
\authorblockN{Yuri Kovajkin}
\authorblockA{
State University of Telecommunication \\
St. Petersburg, Russia  \\
E-mail: viyak@bk.ru}
 }

\maketitle

\begin{abstract}
We develop an approach of key distribution protocol (KDP) proposed recently by T. Aono et al. A more general mathematical model based on the use of Variable-Directional Antenna (VDA) under the condition of multipath wave propagation is proposed. Statistical characteristics of VDA were investigated by simulation, that allows us to specify model parameters. The security of the considered KDP is estimated in terms of Shannon's information leaking to an eavesdropper depending on the mutual locations of the legal users and the eavesdropper.

Antenna diversity is proposed as a mean to enhance the KDP security. In order to provide a better agreement of the shared keys it is investigated the use of error-correcting codes.
\end{abstract}

\IEEEpeerreviewmaketitle

\section{Introduction}

The problem of {\em key distribution} is still in focus of research activity especially for wireless LAN systems. This is due to the severe restriction of asymmetric (public key) cryptography WLAN implementation entailing a lower processing speed.

In order to solve this problem, quantum cryptography~\cite{i} which allows eavesdropping detection within the key sharing procedure seems useful. However, this approach does not reach a practical level due to many technical problems, such as the requirement of special quantum devices.
There are well known {\em key distribution protocols} (KDP) based on the presence of noise in both legal and illegal channels~\cite{ii,iii,iv}. But even though the eavesdropper's channel is less noisy than the legal ones and the eavesdroppers is passive, it is necessary to have the knowledge of the eavesdropper's noisy power in order to guarantee a fixed level of key security. Unfortunately this condition cannot be taken for granted because an eavesdropper may be able to get some advantage at the cost of better receiver sensitivity, or a shorter distance of interception that it was considered by legal parties in the design of the secure KDP.

The most basic assumption on the executed KDP is that the legal and illegal users have different locations, and this fact has to be verified by physical means. (For that matter, an existing special zone surrounding each legal user shall be assumed where the presence of an eavesdropper is not allowed.)

If it is wanted to share a secret key by wireless communication among legal users, it is necessary that one user generates some randomness and then to transmit it to its correspondent in such a way that it is effectively delivered to the legal recipient and any eavesdropper perceives either uncorrelated or weak correlated randomness. 

It is possible to provide non-unit correlation under the condition of {\em multipatch wave propagation}. Let us consider the following mathematical model of the channels between a source of randomness (the first legal user) and both the second legal user and the eavesdropper:
$\eta = \sum_{i=1}^mx_i\xi_i$, $\zeta = \sum_{i=1}^my_i\xi_i$, 
where $\xi=\left(\xi_i\right)_{i=1}^m$ is the vector randomness, $x=\left(x_i\right)_{i=1}^m$ is the coefficient vector of the multipath propagation to the second legal user, and $y=\left(y_i\right)_{i=1}^m$ is the coefficient vector of multipath propagation to the eavesdropper.
Let us assume for simplicity $E(\xi)=0$, then the following relation for the correlation coefficient of $\eta$ and $\zeta$ results:
$$
\rho(\eta,\zeta) = \frac{x^TR_{\xi}y}{\sqrt{(x^TR_{\xi}x)(y^TR_{\xi}y)}}, 
$$
where $R_{\xi}$ is correlation matrix of the random vector $\xi$.

In a general case $\rho(\eta,\zeta)\leq 1$. Moreover if $x$ and $y$ are orthogonal, (e.g. $\langle x,y\rangle=0$) and $R_{\xi}=\mbox{Id}_m$ , then $\rho(\eta,\zeta)=0$.

Common randomness results from fluctuation of the cannel characteristics due to communicating object motion. Such approach has been proposed in~\cite{vii,v}. But it still entails another problem: it is easy to break the secret key under an environment with small fluctuation of the channel characteristics or in the case when the communicating objects are stopped.
In order to overcome these defects, a more sophisticated method, using smart antenna excited randomly by electronic means~\cite{vi}, has been proposed. However, the results presented in this paper were obtained experimentally and the investigation of KDP security performed incompletely is extended here.

The goal of the current paper is thus to introduce  a mathematical model and to present a theoretical investigation concerned with KDP security and reliability based on the use of a {\em Variable-Directional Antenna} (VDA). In order to justify the statistical characteristics of the VDA, we perform a simulation of a ring type VDA that is also excited randomly. In Section~\ref{sc.two}, we describe the conditions of the physical channel and we introduce an exact mathematical model of the KDP. The results of the VDA simulation are presented in Section~\ref{sc.three}. Section~\ref{sc.four} contains an optimization of the KDP in order to provide both reliability and security. Finally we conclude the main results and present some open problems in Section~\ref{sc.five}.

\section{KDP based on multipatch wave propagation and randomly excited VDA}\label{sc.two}

\begin{figure}[!t]
\centering
\includegraphics[width=3in]{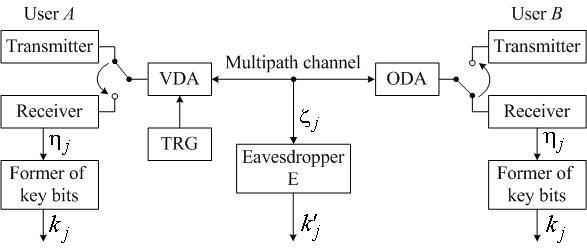} \hspace{1em} 
\caption{Scheme of the communication system corresponding to the KDP.}
\label{fig.01}
\end{figure}

The scheme of the communication system corresponding to the KDP is presented in Fig.~\ref{fig.01}.

The KDP is described in the following steps:
\begin{enumerate}
 \item The legal user A forms the random antenna diagram by exciting the VDA with output of {\em truly random generator} (TRG) and fix this diagram for some given time interval $[0,T_j]$ of the $j$-th key bit generation, $j=1,2,\ldots ,n$.
 \item A transmits to B a harmonic signal $S_j(t)=\cos\omega_0t$, $0\leq t\leq T_j/2$, with the beam pattern obtained at step 1 over the multipath channel.
 \item B receives a harmonic signal from an omni-directional antenna (ODA) and forms the $j$-th key bit by comparing some functional computed with the received signal on the time $[0,T_j/2]$ with a given threshold.
 \item The user B switches off its ODA in a regime of radiation and transmits the same harmonic signal $S_j(t)=\cos\omega_0t$ within the time interval $T_j/2\leq t\leq  T_j$.
 \item The user A switches off its VDA to a receiver and processes the received signal in the same manner as B did, forming the $j$-th key bit.
 \item A and B repeat $n$  times the steps 1--5 with new and independent outputs of TRG in order to create the desired number of key bits.
\end{enumerate}
Thanks to the {\em Reciprocity Theorem} of radio wave propagation between uplink and downlink, the key sequences of A and B should be identical up to a random noise of receivers. Then the signal received by B at time $T_j/2$ can be expressed as:
\begin{equation}
y_j(t) = \sum_{i=1}^m\upsilon_{ij}\beta_{ij}\cos(\omega_0t + \theta_{ij}), \label{eq.03}
\end{equation}
where, with respect to the $i$-th ray at the $j$-th time interval, $\beta_{ij}$ is the channel attenuation coefficient,
$\upsilon_{ij}$ is the VDA amplitude gain,
$\theta_{ij}$ is the VDA phase shift, including both phases in antenna diagram and phase shift in $i$-th ray, and
$m$ is the number of paths (rays).

The signal received by E at time $T_j/2$ is:
\begin{equation}
z_j(t) = \sum_{i=1}^m\upsilon_{ij}'\beta_{ij}'\cos(\omega_0t + \theta_{ij}'), \label{eq.04}
\end{equation}
where the primed parameters have the same meaning as the corresponding parameters in~(\ref{eq.03}) but in possession of E.
(We neglect initially the noise at the legal receivers, and we assume at all moment a noise absence at the eavesdropper E, in advantage with the legal users.)

\begin{figure}[!t]
\centering
\includegraphics[width=2.5in]{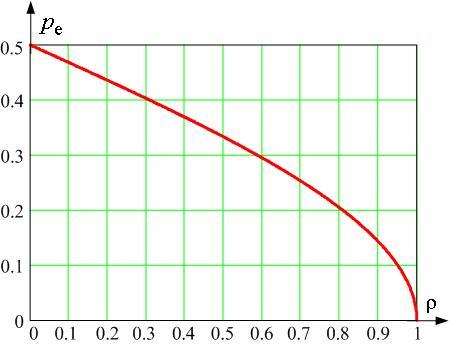} \hspace{1em} 
\caption{The probabilty of the key bit disagreement between legal and illegal users depending on the correlation coefficient $\rho$.}
\label{fig.02}
\end{figure}

Later we will show that the probability distributions of the random values $\eta_j$ and $\zeta_j$,  which are produced by executing some functionals from both $y_j(t)$ and $z_j(t)$  can have a good approximation by a zero mean Gaussian law. Then we prove that the probability of a bit disagreement between the $j$-th bit of the legal users and the eavesdropper key bits obtained by comparing them with a zero threshold is:
\begin{eqnarray}
p_e &=& 2 \int_{-\infty}^0dy \int_0^{+\infty} \frac{\mbox{\rm exp}\left(-\frac{x^2-2\rho xy+y^2}{2\sigma^2(1-\rho^2)}\right)}{2\pi\sigma^2\sqrt{1-\rho^2}}\,dx  \nonumber \\
 &=& \frac{1}{\pi}\mbox{arc\,tan}\left(\frac{\sqrt{1-\rho^2}}{\rho}\right),  \label{eq.05}
\end{eqnarray}
where $\rho$ is the correlation coefficient between $\eta_j$ and $\zeta_j$, $\sigma^2 = \var{\eta_j} = \var{\zeta_j}$.
The dependence of $p_e$ versus $\rho$ is presented in Fig.~\ref{fig.02}. We can see that in contrast to our intuition, the probability $p_e=0.1$ can be provided even when the correlation coefficient $\rho$  has a significant value 0.95.

In order to enhance the security of 
the legal user key string ${\bf k}$ shared after completion of the KDP
it should be performed a privacy amplification~\cite{iii,viii,ix,xii}, or more specifically a mapping of the raw key string ${\bf k}$ to a shorter key string $\tilde{\bf k}$  of length $\ell<n$, using the so called hashing procedure $\tilde{\bf k}=h({\bf k})$  taken from the universal class of hash functions~\cite{x}. Then the amount of Shannon's information leaking to E given her knowledge of the string ${\bf k}'$  satisfies
\begin{equation}
I(\tilde{\bf k};{\bf k}') \leq \frac{1}{2^{n-\ell-t}\ln(2)}, \label{eq.06}
\end{equation}
where $t=n + n \log_2\left(p_e^2 +(1-p_e)^2\right)$  is the Renyi information under the assumption that the errors in the eavesdropper's key bits occur independently due to the independently generated VDA on each of the $j$-th time intervals. Hence in order to select the parameter $\ell$ we should calculate the correlation coefficient $\rho$  depending on the mutual location of the legal user and the eavesdropper, the properties of VDA and the characteristics of the multipath cannel. A solution for this problem will be presented in the next Section.

\section{Correlation between the values $\eta_j$ and $\zeta_j$}\label{sc.three}

\begin{figure}[!t]
\centering
\includegraphics[width=2.5in]{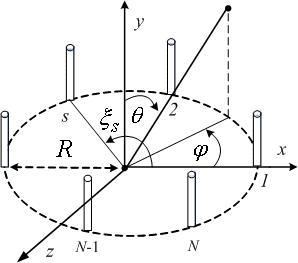} \hspace{1em} 
\caption{Ring antenna with $N$ identical radiators.}
\label{fig.03}
\end{figure}

Let us consider as VDA the so called {\em ring antenna} (RA) shown in Fig.~\ref{fig.03} having $N$ identical isotropic radiators excited by their random phases.

Then the complex instant antenna diagram can be presented by the well known formula~\cite{xi}:
\begin{equation}
f(\phi,\theta) = \sum_{s=1}^N\mbox{\rm exp}\left[ik_0R\sin(\theta)\cos\left(\phi-\frac{2\pi s}{N}\right) - i\psi_s\right],  \label{eq.07}
\end{equation}
where $\psi_s$ is a phase in the $s$-th radiator;
$k_0=\frac{2\pi}{\lambda}$,  $\lambda$ is the length of the wave;
$R$  the radius of the RA; 
$\phi$ is  the angle in the azimuthal plane; and
$\theta$ is  the angle in the vertical plane.

Both instant amplitude and the phase antenna diagrams can be obtained from~(\ref{eq.07}) and they are random values providing random exciting to the RA. It would be possible to find theoretically different statistical characteristics of $f(\phi,\theta)$ but it is rather more easy to solve the same problem by simulation. Since the current paper is limited in space, we present only the main conclusions based on the simulations for the case of independent and uniformly distributed phases $\psi_s$  on $(0,2\pi)$:
\begin{itemize}
 \item the probability distribution of the amplitude antenna diagram has a good approximation through the Rice law which can be approximated in its turn by a Gaussian non-zero mean law;
 \item the probability distribution of the phase antenna diagram has a good approximation by an uniform law on the interval $(0,2\pi)$.
\end{itemize}

\begin{figure}[!t]
\centering
\includegraphics[width=2.5in]{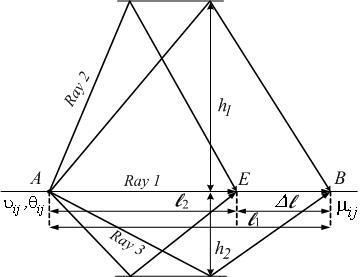} \hspace{1em} 
\caption{Channel model with 3-ray wave propagation.}
\label{fig.04}
\end{figure}

Next it is possible to compute theoretically the correlation coefficients between $\eta_j$ and $\zeta_j$ for different functionals producing them and to find their probability distributions by simulation. However, it is necessary to specify the channel model and the functional description.
To be more specific, let us consider a 3-ray channel model and a location of eavesdropper on the line connecting legal users (Fig.~\ref{fig.04}).

We select two functionals of $y_j(t)$ and $z_j(t)$ producing $\eta_j$ and $\zeta_j$ and the functionals are compared with some thresholds in order to obtain the key  bit $k_j$. The functionals are (see eq.~(\ref{eq.03})):
\begin{itemize}
 \item {\em envelope}: 
$
\mu_j = \sqrt{\mu_{c_j}^2 + \mu_{s_j}^2}  
$ 
where $\mu_{c_j} = \sum_{i=1}^m A_{ij}\cos\theta_{ij}$, $\mu_{s_j} = \sum_{i=1}^m A_{ij}\sin\theta_{ij}$, $A_{ij} = \upsilon_{ij} \beta_{ij}$,
 \item {\em phase difference}
$$
\Delta\psi_j = \psi_{j+1} - \psi_j = \mbox{arc\,tan}\frac{\mu_{s_{j+1}}}{\mu_{c_{j+1}}} - \mbox{arc\,tan}\frac{\mu_{s_j}}{\mu_{c_j}}.  
$$
\end{itemize}
In a similar manner, there can be presented the corresponding functionals for eavesdropper: $\mu_j',\mu_{c_j}',\mu_{s_j}',\Delta\psi_j'$.

We will be interested in finding the probability distributions of all functionals and correlations between similar functionals of any legal user B and the eavesdropper E. Because it is very hard to compute these values theoretically, we will find them by simulation for some given channel parameters.

\begin{figure}[!t]
\centering
\includegraphics[width=3in]{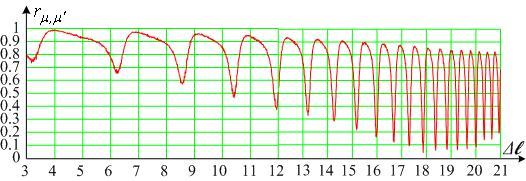} \hspace{1em} \\
(a) \\
\includegraphics[width=3in]{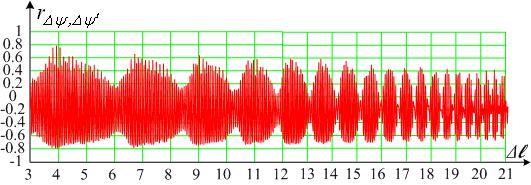} \hspace{1em} \\
(b) \\
\caption{The dependence of correlation coefficients versus distances between legal user and eavesdropper. a) for envelope, b) for phase difference.}
\label{fig.05}
\end{figure}

Let us take $\ell_1=25m$; $h_1=3m$, $h_2=3m$ (distances to the first and to the second reflecting surfaces, respectively), $N=6$, $\lambda =12.5cm$, $R=\lambda/2$ (see Fig's.~\ref{fig.03} and~\ref{fig.04}). Assume that E is placed between legal users A and B within the interval (3--22)m. The results of simulation are presented in Fig.~\ref{fig.05}. The dependences of the correlation coefficients $r_{\mu,\mu'}$ and $r_{\Delta\psi,\Delta\psi'}$  versus distance $\Delta\ell$ between the eavesdropper E and the legal user B are shown in Fig.~\ref{fig.05}(a) and Fig.~\ref{fig.05}(b). 

Since the correlation between the values $\Delta\psi_j$  and $\Delta\psi_j'$  occurs less than the correlation between $\mu_j$  and $\mu_j'$ (see  Fig.~\ref{fig.05}),  it is reasonable to select the phase difference functional in order to form $\eta_j$ and compare it with zero threshold for the $k_j$ key bit generation. (In order to coincide phases of support generators at users A and B, it is possible to transmit a special pilot signal and to tune phases of both users at the initial stage of KDP.)

\begin{figure}[!t]
\centering
\includegraphics[width=3in]{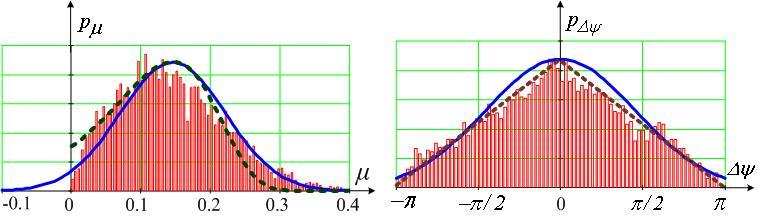} \hspace{1em} \\
 \scriptsize{(a) envelope} \hspace{2cm}  \scriptsize{(b) phase difference}
\caption{Empirical probability distribution for chosen functionals.}
\label{fig.06}
\end{figure}

In Fig.~\ref{fig.06} there are presented empirical probability distributions for these functionals. It is evident that both cases can be approximated by appropriated Gaussian distributions (see solid curves). Therefore the relation~(\ref{eq.05}) can be used to find the probability of disagreement between the key bits of the legal users and the eavesdropper.
But before we address to eq.~(\ref{eq.06}) in order to calculate security of KDP, it should be taken into account an opportunity for the presence of noise at the receivers of the legal users. 

\section{KDP optimization under noisy legal channel}\label{sc.four}

From now on we remove our previous assumption that the multipath channel among legal users A and B is noiseless but keep such condition for eavesdropper's channel. (Obviously, the last assumption cannot degrade the security of KDP.)

In this setting it is necessary to use some methods in order to correct disagreements in key bits of legal users. It is very reasonable to use firstly a selection of the most reliable key bits with a public discussion over a noiseless channel between legal users, and then to apply {\em forward error correction codes} (FEC) by sending of the check bits over the same noiseless channel. (It is worth to note that a noiseless public channel among legal users can be arranged by the choice of special regime, namely large signal power or omnidirectional antenna of the user A that we were unable to use for the execution of KDP.)
The first method of the most reliable key bit selection is to take decision following the rule:
$$
k_j = \left\lbrace \begin{array}{cl}
1 & \mbox{if }\eta_j\geq\alpha, \\
0 & \mbox{if }\eta_j\leq-\alpha, \\
\mbox{erase} & \mbox{otherwise, }
\end{array} \right. 
$$
where $\eta_j$ is the output of $\Delta\psi_j$, and $\alpha$ a threshold.

After a completion of the KDP including a production of the erased bits for both legal users it is necessary to mutually announce the numbers of these bits over public noiseless channels.
In this case, the probability of a key bit disagreement between legal users and eavesdropper, given by~(\ref{eq.05}), has to be corrected because an eavesdropper is able to intercept information about the numbers of accepted key bits over the public channel. We will take into account this fact later for the simulation procedure.
The second method is to keep only the most reliable key bits, say $M$, and to remove the others. This means that the legal users form variation series of the values $|\eta_j|$ on a decreasing order and next to keep (after mutual public discussion) the first $M$ members of this series to generate the key bits.
Of course in this case the probability of key bit disagreement $p_e$ is changed also against~(\ref{eq.05}). 

Let us denote by $p_1$ and $p_2$ the probability of legal key bit errors after the first and the second method, respectively.
Next we use an error-correcting code $(n_0+r, n_0)$ sending a sequence of $r$ check symbols over public noiseless channel in order to correct eventually errors in the key sequence. Then the probability of erroneous decoding $P_{ed}$ by the modified Gallager's theorem is~\cite{xii}:
$P_{ed}\leq 2^{-n_0\,E(R_C)},$ 
where
\begin{eqnarray*}
E(R_C) &=& \max_{\rho_0\in(0,1)}\left[E_0(\rho) - \frac{\rho_0(2R_C-1)}{R_C}\right], \\
E_0(R) &=& \rho_0 - (1+\rho_0)\log_2\left[p^{\frac{1}{1+\rho_0}} + (1-p)^{\frac{1}{1+\rho_0}}\right], 
\end{eqnarray*}
$R_C = \frac{n_0}{n_0+r},$ and $n_0$ is the number of bits $k_j$ which have been kept by legal users after erasing the unreliable bits following the first or the second procedures, and $p$  is the error probability for the kept bits.
In the case of check symbol sending, the Privacy Amplification Theorem against~(\ref{eq.06}) becomes~\cite{xii}:
$$
I(\tilde{\bf k};{\bf k}') \leq \frac{1}{2^{n_0-\ell-t-r}\ln(2)}. 
$$
{\em KDP optimization problem} is to get the maximum key rate 
$$R_{\bf k} = \frac{\ell}{n_0+n_{er}} = \frac{\ell}{n},$$
while $n_{er}$ is the number of erased symbols after the use of the method 1 or 2 and given the values $I(\tilde{\bf k};{\bf k}')$, $P_{ed}$, $\ell$, and different signal-to-noise ratio (S/N) at the receivers of the legal users. We solve this problem by simulation for the case of Gaussian noise at the legal receivers.

In Tables~\ref{tb.01} and~\ref{tb.02} there are presented the results of such optimization for typical conditions for the first and the second method of unreliable bits removal, respectively, where $P_{er}$ is the probability of key bit erasing.
\begin{table}[!t]
$$\begin{array}{||l|l|l|l|l|r|r|l||} \hline  \hline
\rho & \alpha_{opt}&p_e&P_{er}& p_1     & \ell&  n_0  & R_{\bf k} \\  \hline  \hline
     &      &        &        &         & 128 &  9300 & 0.014 \\ 
0.99 & 0.1  & 0.027  & 0.1132 & 0.0032  & 256 & 13100 & 0.02  \\ 
     &      &        &        &         & 512 & 20600 & 0.025 \\  \hline
     &      &        &        &         & 128 &  1200 & 0.107 \\ 
0.95 & 0.1  & 0.089  & 0.1132 & 0.0032  & 256 &  1950 & 0.131 \\ 
     &      &        &        &         & 512 &  3420 & 0.15  \\  \hline
     &      &        &        &         & 128 &   350 & 0.37  \\ 
0.8  & 0.15 & 0.24   & 0.1438 & 0.00035 & 256 &   580 & 0.44  \\ 
     &      &        &        &         & 512 &  1055 & 0.49  \\  \hline \hline
\end{array}$$
\caption{Key rate maximization for the first method given $I(\tilde{\bf k};{\bf k}')=10^{-9}$ bit,
$P_{ed}=10^{-5}$, S/N=100 and different $\rho$.}
\label{tb.01}
\end{table} 

\begin{table}[!t]
$$\begin{array}{||l|l|l|l|l|r|r|l||} \hline  \hline
\rho & M_{opt}&p_e&P_{er}& p_2     & \ell&  n_0  & R_{\bf k} \\  \hline  \hline
     &      &        &        &         & 128 &  7500 & 0.017 \\
0.99 & 9000 & 0.019  & 0.15   & 0.0008  & 256 & 12100 & 0.021 \\
     &      &        &        &         & 512 & 21300 & 0.024 \\  \hline
     &      &        &        &         & 128 &  1130 & 0.113 \\
0.95 & 9000 & 0.084  & 0.15   & 0.0008  & 256 &  1820 & 0.141 \\
     &      &        &        &         & 512 &  3200 & 0.16  \\  \hline
     &      &        &        &         & 128 &   360 & 0.36  \\
0.8  & 9000 & 0.24   & 0.15   & 0.0008  & 256 &   605 & 0.42  \\
     &      &        &        &         & 512 &  1090 & 0.47  \\  \hline  \hline
\end{array}$$
\caption{Key rate maximization for the second method given $I(\tilde{\bf k};{\bf k}')=10^{-9}$ bit,
$P_{ed}=10^{-5}$, S/N=100 and different $\rho$.}
\label{tb.02}
\end{table} 

We can see from these tables that the second method is for large correlation a little bit better than the first one. However both methods provide sufficiently reliable and secure key sharing if eavesdropper is placed on 3--21m away from legal user B and phase difference is used as key generating method (see Fig.~\ref{fig.05}(b)). A similar conclusion is drawn also for multipath channels with other parameters and locations of eavesdroppers.
In order to enhance the security of the KDP, antenna diversity can be used when B has $m$ omnidirectional antennas and he selects randomly one of them at each time period $T_j$ to receive and transmit signal. Then the relation finding the Renyi information used in~(\ref{eq.06}) changes for:
\begin{equation}
t=n + \frac{n}{m} \log_2\left(\tilde{p}_e^2 +(1-\tilde{p}_e)^2\right). \label{eq.13}
\end{equation}
The relation~(\ref{eq.13}) holds with the probability equal to the probability of the event in which with at least of one of antennas a mutual location of the legal user and the eavesdropper is got such that $\rho\leq\rho^*$, where $\rho^*$  is found by~(\ref{eq.05}) given $\tilde{p}_e$.

We considered so far a scenario when an eavesdropper uses the same omnidirectional antenna as the legal user B. But E can execute directional antenna to separate all rays and to process the best of them or even apply joint processing to all of them. We have performed a simulation of the case with single ray separation and it has been shown that the correlation coefficient even decreases in comparison with one presented before. The case of joint processing of separated rays is noteworthy. But we can remark that even under the very strong condition in which the eavesdropper knows exactly all channel parameters both for E and B, there is still uncertainty about VDA gains in the direction of E and B. Therefore, generally speaking, the correlation coefficient occurs even in this case with a value less than one.

\section{Conclusion and future work}\label{sc.five}

We considered a method of key sharing based on the concept of a VDA under the condition of multipath channel and we showed that sufficient security and reliability of the shared keys can be provided even when the eavesdropper's channel is noiseless.
The results of investigations show that the security of the KDP (in terms of Shannon's information leaking to eavesdropper) does not depend only on the distance between legal users and eavesdropper but also on the eavesdropper's location. This result somewhat contradicts to a very optimistic conclusion in~\cite{vi}.

We propose to use the difference-phase functional instead of either quadrature components or envelope in order to form key bits. This approach results in less mutual correlation between legal user and eavesdropper and simplifies a choice of threshold. The key sequence ${\bf k}$ is i.i.d if VDA is excited by independent random phases and threshold is chosen in an appropriate manner. (This fact has been confirmed by simulation using statistical tests.)
Our contribution consists also in the proof of relation~(\ref{eq.05}) which allows to connect the probability of disagreement between the key bits of legal users and eavesdropper with the correlation of corresponding values.
Unfortunately, a limited space of the paper does not allow us to show all simulation results for different multipath channels and mutual location of legal users and eavesdroppers, which we have got at our disposition.

In the future we are going to investigate: i) the use of multitone signals in the KDP, ii) the localization of optimal processing of the eavesdropper rays separation  in order to provide the greatest correlation, iii) the use of real FEC and effective decoding algorithms with KDP (instead of extended Gallager's bounds); and, iiv) the use of other types of VDA (like ESPAR or others).

\end{document}